\documentclass[fleqn,twoside]{article}
\usepackage{espcrc2}

\usepackage{graphicx}
\usepackage[figuresright]{rotating}

\newcommand{\AmS}{{\protect\the\textfont2
  A\kern-.1667em\lower.5ex\hbox{M}\kern-.125emS}}

\title{Test of the Polyakov Loop Model}

\author{Adrian Dumitru and Robert D. Pisarski
\address{Department of Physics\\
	Brookhaven National Laboratory\\
	Upton, NY 11973 USA\\}
        \thanks{This research was supported by
	DOE grant DE-AC-02-98CH-10886.}}
       
\begin{document}

\begin{abstract}
We discuss the two point functions for the real and imaginary parts
of the Polyakov loop in a pure
$SU(3)$ gauge theory.  The behavior of these correlation functions 
in the Polyakov Loop Model is markedly different
from that in perturbation theory.
\vspace{1pc}
\end{abstract}

\maketitle

Consider the behavior of an $SU(3)$ gauge theory without
dynamical quarks.  The usual quantity measured is the
pressure, as a function of the temperature.  While important, and
indeed the only thing which one needs for thermodynamics, there
are many other things to measure, such as
the correlation functions of gauge-invariant operators.

In this note we discuss what certain correlation functions may
tell us about the behavior of the deconfined phase.  We work
within the context of the Polyakov Loop Model 
\cite{rp1,wirstam,ogilvie}.  For reasons which will become
clear later, however, it is probably imperative to think
of how to parametrize these correlation functions in a
manner {\it in}dependent of any theoretical prejudice.

At a nonzero temperature $T$, a fundamental quantity is the
thermal Wilson Line,
\begin{equation}
{\bf L}(\vec{x}) \; = \; 
{\cal P} \exp \left( i g \int^{1/T}_0 A_0(\vec{x},\tau) \, 
d\tau \right) \; .
\end{equation}
This transforms as an adjoint field under the local $SU(3)/Z(3)$ 
gauge symmetry, and as a field with charge one under the global
$Z(3)$ symmetry.  To obtain a gauge invariant operator, the
simplest thing to do is to take the trace, forming the Polyakov
loop,
\begin{equation}
\ell_1 = \frac{1}{3} \; {\rm tr}\left( {\bf L} \right) \; .
\end{equation}
This transforms under the global $Z(3)$ as a field with charge one.
The expectation value of $\ell_1$ is only nonzero above $T_c$,
which is the temperature for the deconfining phase transition.

In fact $\ell_1$ is only the first in an infinite series of 
gauge-invariant operators.  For example, consider 
\begin{equation}
\ell_2 = \frac{1}{3} {\rm tr}\left( {\bf L}^2 \right) - \ell_1^2 \; .
\end{equation}
Under the global $Z(3)$ symmetry,
this Polyakov loop has charge two.  The charge one part
of ${\rm tr}( {\bf L}^2)$, $\ell_1^2$, is subtracted off
to obtain an independent field.
In this note we concentrate only upon the charge one Polyakov loop,
$\ell \equiv \ell_1$, and drop all Polyakov loops with other charges,
such as $\ell_2$, the singlet field 
$\ell_3 \sim {\rm tr}({\bf L}^3) + \ldots$, {\it etc.}  

We begin with a potential for the Polyakov loop, taking
the simplest form consistent with the 
global $Z(3)$ symmetry:
\begin{equation}
{\cal V}(\ell) = - \frac{b_2}{2} |\ell|^2 - \frac{b_3}{3}
\left( \ell^3 + (\ell^*)^3 \right) + \frac{1}{4}\left( |\ell|\right)^2 \;
\end{equation}
The coefficient of the quartic term is chosen to simplify further
results.  At the minimum of the potential, which we assume
occurs for real $\ell$, $\partial{\cal V}/\partial \ell = 0$, 
\begin{equation}
\ell_0 \equiv \langle \ell \rangle = b_3 + \sqrt{b_2 + b_3^2} \; .
\end{equation}
In the Polyakov Loop Model, 
the pressure is related to the potential as:
\begin{equation}
p(T) = - {\cal V}(\ell_0) b_4 T^4 \; .
\end{equation}
At high temperatures, $b_2$ is adjusted so
that $\ell_0 \approx 1$; then $b_4$ is
adjusted to give the proper value of the ideal gas term.
Away from infinite temperature, 
in the spirit of mean field we take the quantities $b_3$
and $b_4$ to be approximately constant with temperature.  
Given the pressure, the dependence of $b_2$ upon the
temperature is then fixed.

While $\ell_0$ is the standard variable measured on the lattice,
this is the bare value.  
Single insertions of the Polyakov loop are regularized by
introducing a renormalization constant;
the natural condition to fix the value of that constant
is to require that the renormalized Polyakov loop is unity at infinite
temperature \cite{yaffe}.  If a lattice regulator is used instead of
dimensional regularization, though, one has to deal with divergences
$\sim g^2/(a T)$, {\it etc.}, which are most singular as the lattice spacing
$a \rightarrow 0$.  

Thus at present,
we cannot easily relate the one point function of the Polyakov
loop, as measured on the lattice, to the pressure.  However,
we now show that for three colors, one can relate certain two
point functions of the Polyakov loop, to the pressure, in an
unambiguous fashion.

For $SU(3)$, the Polyakov loop is a complex number,
with real, $\ell_r = {\rm Re} \ell$, and imaginary,
$\ell_i = {\rm Im} \ell$, parts.  
By a global $Z(3)$ rotation, we can assume
that the vacuum expectation value of $\ell$, $\ell_0$, is real.  
Computing second derivatives, the mass squared for
the real part is:
\begin{equation}
m^2_r = \frac{\partial^2 {\cal V}}{\partial \ell_r^2}
= - b_2 - 4 b_3 \ell_0 + 3 \ell_0^2 \; ,
\end{equation}
while that for the imaginary part is:
\begin{equation}
m^2_i = \frac{\partial^2 {\cal V} }{\partial \ell_i^2}
= - b_2 + 4 b_3 \ell_0 + \ell_0^2 \; .
\end{equation}

When $b_3 \neq 0$, the transition is necessarily of first order.
The transition occurs when the nontrivial minimum
is degenerate with the trivial minimum; {\it i.e.}, when 
${\cal V}(\ell_0) = 0$.  Putting in the expression for
$\ell_0$, we find
\begin{equation}
b_2(T_c^+) = - \frac{8}{9} b_3^2 \;\;\; , \;\;\; 
\ell_0(T_c^+) = \frac{4}{3} b_3 \; .
\end{equation}

This is all trivial algebra, done in detail to avoid any
possible confusion.  The full effective lagrangian can
then be computed.  Besides
the potential term, given above, there is also the kinetic term,
with a nonstandard normalization:
\begin{equation}
{\cal Z}_W T^2 |\vec{\partial} \ell|^2 \;\;\;, \;\;\;
{\cal Z}_W = \frac{3}{g^2} \left( 1 - .08 \frac{g^2}{4 \pi} 
+\ldots \right) \; .
\end{equation}
The first term in ${\cal Z}_W$
appears at the classical level, while the second arises from
one loop corrections, as computed by Wirstam \cite{wirstam}. 

Over large distances, 
$x\rightarrow \infty$,
the two point functions of the Polyakov loop are
\begin{equation}
\langle \ell_{r}(x) \ell_{r}(0) \rangle - \langle \ell \rangle^2 
\sim \frac{\exp(- \widetilde{m}_{r} x)}{x} \;\;\; 
\; ,
\end{equation}
\begin{equation}
\langle \ell_{i}(x) \ell_{i}(0) \rangle \sim 
\frac{\exp(- \widetilde{m}_{i} x)}{x} \;.
\end{equation}
The two fields, $\ell_r$ and $\ell_i$, don't mix to the order at which
we work. The masses which enter into the correlation functions are
\begin{equation}
\widetilde{m}_{r,i}^2 = \frac{b_4}{{\cal Z}_W} m_{r,i}^2 T^2 \; .
\end{equation}

For two colors, the Polyakov loop is real, and one can only measure
one mass.  Then, without knowing both the coupling constant
and the wave function renormalization constant ${\cal Z}_W$, 
there is no firm prediction.

This is not true for three colors.  Then one can form the {\it ratio} of
the masses for the real and imaginary parts of the Polyakov loop.
The constants $b_4$ and ${\cal Z}_W$ drop out, and one has a unique
relation between this ratio of masses and the pressure.  
In particular, at the point of transition, using the previous results
we find that
\begin{equation}
\frac{\widetilde{m}_i}{\widetilde{m}_r} = 3 \;\;\; , \;\;\;
T=T_c^+ \; .
\label{e1}
\end{equation}
This is our principal result.  It is dependent upon the
assumed form of the potential for ${\cal V}(\ell)$, and would
change if terms such as $\sim (|\ell|^2)^3$ were included.

These two point functions in
the Polyakov Loop Model are very different from
those of ordinary perturbation 
theory.  In perturbation theory, $\ell_0$ is near unity, and
correlations are determined by multiple exchanges of $A_0$ fields.
The mass of the $A_0$ field is
the Debye mass, $m^2_D \sim g^2 T^2$.  Expanding the exponentials,
the real part of the Polyakov loop couples to $\sim {\rm tr}A_0^2$, while
that for 
the imaginary part couples to $\sim {\rm tr}A_0^3$.  Thus over large distances,
$x \rightarrow \infty$, 
\begin{equation}
\langle \ell_{r}(x) \ell_{r}(0) \rangle - \langle \ell \rangle^2 \sim 
\frac{\exp(- 2 m_D x)}{x^2} \; ,
\end{equation}
\begin{equation}
\langle \ell_{i}(x) \ell_{i}(0) \rangle \sim 
\frac{\exp(- 3 m_D x)}{x^3} \; .
\end{equation}
Notice that the prefactors in front differ markedly from those
of the Polyakov Loop Model; instead of $1/x$, they are $1/x^2$
and $1/x^3$, respectively, with the power of $1/x$ measuring the
number of quanta exchanged.

If we ignore the difference in prefactors, even so the perturbative
result for the mass ratio of (\ref{e1}) is not $3$, but $3/2$.

Measurements of the two point function of the real part of the Polyakov
Loop have been carried out by Kaczmarek {\it et al.} \cite{potential}.
From the two point function of Polyakov loops, which is 
presumably dominated
by that for the real part, the mass drops by about a factor
of ten, from $m/T \sim 2.5$ at $T=2 T_c$, to perhaps
$m/T \sim .25$ at $T_c^+$.  We are not aware of any measurements
of the imaginary part close to $T_c$.  

There are also measurements by Bialas {\it et al.} for a $SU(3)$ gauge
theory in $2+1$ dimensions \cite{pi}.  While the critical behavior in this
model is that of a two dimensional system, and so can have characteristics
special to a low dimension, for the Polyakov Loop Model 
in mean field theory, our predictions remain the same.  These
authors find that the ratio $\widetilde{m}_i/\widetilde{m}_r$ does
increase from $3/2$ as the temperature approaches $T_c$.

In fact, the Polyakov Loop Model must be inapplicable at some 
temperature not too far above $T_c$.  At high temperature, where
$\ell_0 \approx 1$, the above formula give $b_2 = 1 - 2 b_3$, and
\begin{equation}
\frac{\widetilde{m}_i}{\widetilde{m}_r} = 
\sqrt{ \frac{3 b_3}{1 - b_3} } \;\;\; , \;\;\;
T \rightarrow \infty \; .
\end{equation}
The constant $b_3$ is not well determined, but for $b_3 < 3/7$, the
above ratio is less than the perturbative value of $3/2$.

Thus we propose that the two point function of Polyakov loops can
be used as a measure of the regime in which the Polyakov Loop Model
applies, and the regime where perturbation theory applies.

What if the ratio of masses, (\ref{e1}), is wrong even at $T_c$?  Besides
including terms of higher order in the potential, it may also be
necessary to include the charge two Polyakov loop,
$\ell_2$.  Since $Z(3)$ is a cyclic symmetry, a field with
charge two is the same as one with charge minus one.
The couplings
of this loop with itself are identical to the couplings of the
usual Polyakov loop, since the sign of the charge doesn't matter.
Unlike the charge one Polyakov loop, however, the charge two loop
should always have a positive mass squared, in order to avoid 
condensation which breaks $SU(3) \rightarrow SU(2)$ \cite{rp1}.
Thus one would assume that the charge two field, as a massive
field, can be ignored.  Nevertheless, the following coupling
is $Z(3)$ symmetric:
\begin{equation}
\ell_1 \ell_2 + \ell_1^* \ell_2^*
\end{equation}
This term mixes the charge one and
charge two Polyakov loops, 
$\sim {\rm tr}{\bf L} {\rm tr}{\bf L}^2  + \ldots$. 
Its coupling constant is directly measurable; if small,
the charge two Polyakov loop can 
be ignored, and our prediction should hold.

Lastly, we note that the Polyakov loop may well couple weakly to
other operators.  Thus while in principle it should dominate all
correlation functions at large distances (if it is the lightest
state), this may be very difficult to see unless the operator
couples strongly.


\begin{thebibliography}{9}
\bibitem{rp1} R. D. Pisarski, Phys. Rev. D62 (2000) 111501;
A. Dumitru and R. D. Pisarski, Phys. Lett. B504 (2001) 282; hep-ph/0106176.
\bibitem{wirstam} J. Wirstam, hep-ph/0106141.
\bibitem{ogilvie} P. N. Meisinger, T. R. Miller, and M. C. Ogilvie,
hep-ph/0108009; P. N. Meisinger and M. C. Ogilvie, hep-ph/0108026.
\bibitem{yaffe} L. G. Yaffe, private communication.
\bibitem{potential}
O. Kaczmarek, F. Karsch, E. Laermann, and M. Lutgemeier,
Phys.Rev. D62 (2000) 034021.
\bibitem{pi}
P. Bialas, A. Morel, B. Petersson, K. Petrov, and T. Reisz, these
proceedings.
\end{thebibliography}
\end{document}